\documentstyle[prd,preprint,aps,epsfig,axodraw]{revtex} 
\begin{document} 
 
%\baselineskip 1.2 \baselineskip 
 
%\draft 
 
\newcommand{\TeV}{\,{\rm TeV}} 
\newcommand{\GeV}{\,{\rm GeV}} 
\newcommand{\MeV}{\,{\rm MeV}} 
\newcommand{\keV}{\,{\rm keV}} 
\newcommand{\eV}{\,{\rm eV}} 
\def\ap{\approx} 
\def\bea{\begin{eqnarray}} 
\def\eea{\end{eqnarray}} 
\def\bi{\begin{itemize}} 
\def\ei{\end{itemize}} 
\def\be{\begin{enumerate}} 
\def\ee{\end{enumerate}} 
\def\ler{\lesssim} 
\def\gtr{\gtrsim} 
\def\beq{\begin{equation}} 
\def\eeq{\end{equation}} 
\def\haf{\frac{1}{2}} 
\def\nn{\nonumber \\} 
\def\p{\prime} 
\def\pa{\partial} 
\def\ccg{\cal G} 
\def\L {{\cal L}} 
\def\O{\cal O} 
\def\R{\cal R} 
\def\U{\cal U} 
\def\V{\cal V} 
\def\W{\cal W} 
\def\a{\alpha} 
\def\e{\varepsilon} 
\def\rt{\sqrt{2}} 
\def\dy{\Delta y} 
\def\ra{\rightarrow} 
\def\ia#1{\frac{2\pi}{\alpha_{#1}}} 
\def\slash#1{#1\!\!\!\!\!/} 
 
%%%%%%%%%%%%%%%%%%%%% Main Text %%%%%%%%%%%%%%%% 
\setcounter{page}{1} 
%\renewcommend{\arraystretch}{1.3} 
%\draft 
%\widetext 
\preprint{KIAS-P02039, IFT-02/31 \\} 
 
\title{\Large \bf Running of Gauge Couplings in $AdS_5$  \\ via Deconstruction} 
 
\author{$^{(a)}$Adam Falkowski\footnote{\tt Adam.Falkowski@fuw.edu.pl} 
and $^{(b)}$Hyung Do Kim\footnote{\tt hdkim@kias.re.kr} \\} 
\address{\vspace{5mm} 
$^{(a)}$  Institute of Theoretical Physics, Warsaw University \\ 
 Ho\.za 69, 00-681 Warsaw, Poland \\ 
\vspace{2mm} 
$^{(b)} $Department of Physics, Korea Institute for Advanced 
Study\\   Seoul, 130-012, Korea \\} 
\date{\today} 
\tighten 
\maketitle 
 
\begin{abstract} 
Running of gauge couplings on a slice of $AdS_5$ is examined 
using the deconstruction setup. 
Logarithmic running instead of (linear) power law is justified 
when the cutoff is lower than the curvature scale. 
Most of interesting features including the localization of Kaluza-Klein modes and the widening of higher Kaluza-Klein spectrum spacing are well captured 
within the framework of deconstruction. 
\end{abstract}

\newpage 
%%%%%%%%%%%%%%%%%%%%%%%%%%%%%%%%%%%%%%%%%%%%%%%%%%%%% 
\section{Introduction} 
%%%%%%%%%%%%%%%%%%%%%%%%%%%%%%%%%%%%%%%%%%%%%%%%%%%%%%% 
 
  In recent years new solutions to the hierarchy problem based on the presence of extra space-time dimensions have been  proposed. In Ref. \cite{Arkani-Hamed:1998rs,Antoniadis:1998ig} 
 the weakness of gravity compared to the gauge interactions was explained 
with the aid of large volume of flat extra dimensions 
in which only gravitons propagate. 
On the other hand, if the space is curved along the extra dimension, 
one  can achieve a large hierarchy even for a small 
size of the extra dimension \cite{Randall:1999ee}. 
The setup proposed by Randall and Sundrum involved a slice of $AdS_5$ 
space-time terminated by two branes: the Planck brane and the TeV brane where 
the Standard Model fields were assumed to live. 
The hierarchy appears due to  the presence of a large warp factor which 
suppresses all the mass scales (in particular the fundamental (cutoff) scale)  on the TeV brane.  At the same time, 
 the gravitational interactions 
are weak since the massless graviton mode  has a small overlap 
of the wave function  with the TeV brane matter. 
 
Randall and Sundrum  proposed this setup as an  alternative 
to the MSSM or the large extra dimensions. 
 However, the question remains whether the model can be reconciled with 
 the gauge coupling unification  as  the low-energy 
 effective theory breaks down close to TeV energies 
  and quantum gravity effects emerge. There seems no chance 
 to embed in the Randall-Sundrum model  the triumph of 
 gauge  coupling unification in the MSSM since there is no 
huge scale difference which allows three different couplings 
to be unified with large log corrections. 
 
However recently it has been realized that the Randall-Sundrum model can be made compatible with the large-scale unification if the Standard Model gauge fields are allowed to propagate in the bulk $AdS_5$ space-time.  
The question of the running of gauge couplings in 
$AdS_5$ was first studied by Pomarol \cite{Pomarol:2000hp} who showed, 
 that if the cutoff of the 
 theory is below the  curvature scale of $AdS_5$, 
logarithmic running occurs instead of 
linearly divergent power law which happens in flat 5D. 
In Pauli-Villars regularization  all the heavy 
Kaluza-Klein modes are paired up with the modes of the regulating 
ones  and thus they give a negligible contribution for  the running. The large log correction arises due to the 
differences of the infrared(IR) cutoff scale $\mu$ and the ultraviolet(UV) cutoff given by the 
regulator mass. Randall and Schwartz \cite{Randall:2001gb,Randall:2001gc} 
used a different prescription 
inspired by AdS/CFT correspondence in order to show the 
logarithmic running of the bulk gauge couplings. A position dependent 
momentum cutoff was used as a regularization and, 
as a consequence, the gauge coupling got logarithmic corrections 
as long as the cutoff was below the scale of the curvature. 
Choi, Kim and Kim \cite{Choi:2002wx} 
derived  general running equations 
for the gauge couplings in an orbifold of $AdS_5$ in the 
supersymmetric setup based on holomorphy. 
Dimensional regularization \cite{Goldberger:2002cz,Choi:2002zi} 
shows the momentum dependence ($\log p$), the radius dependence ($\log R$) 
and the divergent structure ($\log \Lambda$ or $\log \mu$) 
separately. 
Following this approach, one can derive all the radius 
dependence and the momentum dependence of the gauge couplings by 
reading the Kahler metric and considering possible threshold 
corrections in the presence of the curvature. 
Radius running \cite{Choi:2002zi} 
is a nice way to reach the coupling defined at UV. 
In \cite{Agashe:2002bx}, 
more detailed analysis was done using the Pauli-Villars regularization 
and it confirmed the results of Pomarol. Holographic evolution of 
gauge couplings was discussed in \cite{Contino:2002kc}. 
%using dimensional regularization. 
 
In this paper we approach the problem using the deconstruction 
setup. In Refs. \cite{Arkani-Hamed:2001ca,Hill:2000mu} it was observed 
 that gauge theories in four dimensions with the gauge group 
$SU(N_c)^N$ and scalars (link-Higgs fields) 
 in bifundamental representations appear to be equivalent 
to a higher dimensional  $SU(N_c)$ theory in the flat background 
{\footnote{Similar models had been studied previously for different reasons, 
see \cite{Georgi:au,Halpern:1975yj,Douglas:1996sw}.}}. 
  The correspondence holds below some deconstruction scale $v$ 
set by the expectation values of the link-Higgs fields 
that break the group $SU(N_c)^N$ to its diagonal subgroup. 
 Below $v$ the spectrum and interactions of 
the four dimensional degrees of freedom  are analogous to that of the 
Kaluza-Klein(KK) modes 
 of gauge bosons propagating in extra dimensions. 
 Above the  scale $v$, a deconstruction model can be considered as a 
 (possibly renormalizable) UV completion 
 of the corresponding extra dimensional theory. Hence, there appeared a new possibility to study the extra dimensional physics using the standard tools of 
 four dimensional Yang-Mills  theories. 
Recently it has been  noticed 
that also gauge fields propagating in a curved background 
can be described by four dimensional gauge theories \cite{Abe:2002rj} 
 (see also  \cite{Sfetsos:2001qb}).  This can be simply achieved using 
the analogous set-up as that of \cite{Arkani-Hamed:2001ca,Hill:2000mu} 
but with non-universal vevs of the link-Higgs fields.  
%(which translates into non-universal lattice spacing). 
Thus,  deconstruction can also help to understand the physics of the 
curved space. 
 
 Deconstruction is used in our paper  as an alternative regularization 
method and is compared to other regularizations. 
We show that the Renormalization Group(RG) running equation is directly 
related to the boundary condition for the gauge 
couplings that we start with, and is highly dependent on the 
UV completion.  We restrict the study to the region 
in which the cutoff scale $\Lambda$ 
(the inverse of the lattice size) is lower than the 
curvature scale k ($\Lambda \ll k$) 
though a more general study is possible with the aid of 
numerical analysis for $\Lambda \sim k$. 
The other limit (the cutoff scale is much higher than the 
curvature scale, $\Lambda \gg k$) 
is very similar to the flat space since we can not feel the 
presence of the curvature for $\Lambda \gg k$. 
 There are two interesting observations for $\Lambda \ll k$. 
Firstly, it is shown that in deconstruction there are less heavy 
states than in the low energy effective description of the Randall-Sundrum(RS) model 
  and the number of KK modes increases  logarithmically 
rather than linearly in energy scale. This makes the running of the gauge 
coupling soften compared to the flat case and the rather mild running instead 
of power law running is expected. 
Secondly, the running of the 
gauge coupling above TeV is not a running of the zero mode but rather a running 
of the part of the gauge coupling existing near Planck brane. This is related to 
the definition of the Wilsonian gauge coupling from which we can find the 
relation with the low energy coupling measured by experiment. The 
natural Wilsonian gauge coupling at $\Lambda$ 
is defined at different scales for different 
positions along the fifth dimension (which is $\Lambda$ for Planck brane 
and is $\Lambda e^{-\pi kR}$ for TeV brane),
 and we can not apply the zero mode running 
formula used in the flat space extra dimension. This reduces the effective 
number of KK modes at a given energy scale from $\log E$ to 1, and we get the 
running formula which is just the same as  the four dimensional one. 
% The gauge coupling we consider is defined only at certain parts 
%of the group space from Planck brane and the actual calculation involves the wave function overlap of 
%each mode over that region. 
% Figure 1 and 2 show the differences 
%of the Wilsonian integrating out procedure for $\Lambda > M_{GUT}$ 
%and $\Lambda < M_{GUT}$. The latter one corresponds to the region that 
%we are interested in the paper. 
 
\section{Setup} 
 
In this section we summarize the correspondence between the 5D gauge theories and the deconstruction models. 
In a warped background with the metric 
$ds^2 = a^2(y) \eta_{\mu\nu} dx^\mu dx^\nu - dy^2$ the 5D Yang-Mills action reads: 
\beq 
\label{eq.fga} 
\delta S_{5D} = \int d^4 x \int_0^{\pi R} dy {\rm Tr} \left (-\frac{1}{2}F_{\mu\nu} F_{\mu\nu} + a^2(y) F_{\mu 5} F_{\mu 5}  \right), 
\eeq 
where  the field strength is  defined as 
$F_{\alpha \beta} = 2 \frac{1}{g_5} \pa_{[\alpha} (g_5 A_{\beta ]})+ i g_5 [A_\alpha, A_\beta]$ and $g_5$ is  the gauge coupling (here we allow $g_5$ to be position dependent). 
One can  discretize the compact fifth dimension into $N$ equally sized intervals $(y_i, y_{i+1})$, with the lattice spacing $\dy \equiv y_{i+1}-y_i = \frac{R}{N}$. The integral over the $y$ coordinate  is traded for the sum, $\int dy \ra \sum_{i=1}^N \dy$, and  the derivates along the fifth dimension are discretized as $\pa_5 A_\mu|_{y_i} \ra \frac{A_{\mu,i+1} - A_{\mu,i}}{\dy}$. 
 
The deconstruction models which mimic the 5D $SU(N_c)$ gauge theories compactified on the orbifold $S_1/Z_2$ involve $N$ copies of $SU(N_c)$ group with 
the gauge couplings $g_i$ ($i=1,\cdots,N$) and link Higgs fields $Q_i$ 
($i=1,\cdots,N-1$). The links are bifundamental under $SU(N_c)_i \times 
SU(N_c)_{i+1}$, thus they are represented by $N_c \times N_c$ complex matrices. The action is: 
\beq 
S_{4D} = \int d^4x \sum_{i=1}^N {\rm Tr} \left ( - \frac {1}{2} F_{\mu\nu,i} F_{\mu\nu,i}   +  2 D_\mu Q_i^\dagger D_\mu Q_i + V(Q_i) \right ) 
\eeq 
The link-Higgs field can be conveniently split as $Q_i = \frac{1}{\rt} (\Sigma_i + i G_i)$.  
%Furthermore, we will often abbreviate $A_{\mu,i} \equiv A_i$. 
 The scalar potential $V(Q_i)$ is necessary go give the link-Higgs fields the desired pattern of vacuum expectation values, but the details of it are not relevant for us. 
 
When the link-Higgs bosons acquire vacuum expectation values, $\langle Q_i \rangle = v_i$, the  full gauge symmetry is broken down to the diagonal subgroup and the usual Higgs mechanism gives masses to the gauge fields corresponding to broken generators. The action then  reads: 
\bea 
\label{eq.dga} 
&S_{4D} = \int d^4x\sum_{i=1}^N {\rm Tr} \left ( - \frac {1}{2} F_{\mu\nu,i} F_{\mu\nu,i} +  Z_{\mu,i}^\dagger Z_{\mu,i}   \right) &\nn& 
Z_{\mu,i} =  \left ( \pa_\mu G_i - i \pa_\mu \Sigma_i + v_i (g_i A_{\mu,i} - g_{i+1} A_{\mu,i+1}) 
+ g_i A_{\mu,i} (\Sigma_i + i G_i) - (\Sigma_i + i G_i)g_{i+1} A_{\mu,i+1}    \right) 
\eea 
In  non-supersymmetric models the fields $\Sigma_i$ have no interpretation in terms of 5D degrees of freedom\footnote{In supersymmetric models they are matched to  the real scalar component of the 5D ${\cal N}=2$ gauge multiplet \cite{Csaki:2001em}.}. Thus they should be given a large mass (by adjusting the  scalar potential $V(Q_i)$ appropriately) and decoupled from the low-energy theory. In the following we assume that this step has been done and neglect $\Sigma$ in our analysis. The remaining terms in $Z_{\mu,i}$ can be arranged as follows: 
\beq 
Z_{\mu,i} =  \left (\pa_\mu G_i + v_i g_i  \frac {g_i A_{\mu,i} - g_{i+1} A_{\mu,i+1}}{g_i} 
+ i g_i [A_{\mu,i}, G_i] - i G_i (g_{i+1} A_{\mu,i+1} -g_i A_{\mu,i})    \right) 
\eeq 
Now we can do the following identification between the degrees of freedom in our deconstruction model  and the 5D gauge theory: 
\bea 
\label{eq.gm} 
g_5(y_i) \frac{1}{\sqrt{\Delta y}} &\ra & g_i \nn 
A_\mu(y_i) \sqrt{\Delta y }& \ra &  A_{\mu,i} \nn 
A_5(y_i) \sqrt{\Delta y }& \ra &  G_i \nn 
 \frac{a(y_i)}{\Delta y }  & \ra & g_i v_i 
% \nn  R \hspace{1cm}  & \sim & \hspace{1cm}  \sum_{i=1}^N \frac{1}{g_i v_i} 
\eea 
The problem at what scale should this matching actually be done will be discussed later on. The correspondence between the 4D part of the field strengths $F_{\mu\nu}^2$ in the two theories is  obvious. The other term in (\ref{eq.dga}) becomes: 
\beq 
Z_{\mu,i}^\dagger Z_{\mu,i} = \dy a(y)^2 \left | \pa_\mu A_5 -   \frac{1}{g_5} \pa_5  (g_5 A_\mu) 
+ i g_5 [A_\mu, A_5] - i\dy A_5 \pa_5  (g_5 A_\mu) \right|^2 |_{y_i} 
\eeq 
 and  we recover  the $F_{\mu 5}^2$ term of the discretized 5D action (\ref{eq.fga}) up to the ${\cal O}(\dy^2)$ correction.  In 5D interpretation this correction can be considered as a higher-derivative term suppresed by the appropriate power of the fundamental scale. Thus, in the deconstruction models,  the role of the fundamental scale is played by the inverse of the lattice spacing, $\Lambda  \sim (\dy)^{-1}$. 
 
For $v_i=v$, $g_i = g$ the deconstruction model describes, at low energies, the same physics as that of the 5D gauge theory in the flat background 
\cite{Arkani-Hamed:2001ca,Hill:2000mu}. 
In this case it is straightforward to diagonalize 
the mass matrix for the gauge bosons and the  spectrum turns out to be 
\bea 
m_n & = & 2 gv \sin (\frac{n\pi}{2N}), ~~~~(n=0,\cdots, N-1). 
\eea 
For $n \ll N$, we obtain the well-known spectrum of Kaluza-Klein modes of a 5D 
gauge theory compactified on $S^1/Z_2$, $m_n= \frac{n}{R}$, with the radius 
$R = \frac{N}{\pi g v}$. 
Thus there are three distinctive energy scales. 
Below the scale $\frac{g v \pi}{N}$ the theory 
is analogous to the ordianary 4D gauge theory. 
In the range from  $\frac{g v \pi}{N}$ up to the 
inverse lattice spacing $g v$ the heavy gauge bosons 
can be seen and the physics is analogous to that of the 5D gauge theory. 
Finally, above  $g v$ the theory is an unbroken 4D gauge theory 
with the product group $SU(N_c)^N$. 
In this energy range the deconstruction model can be considered 
as one of many possible UV completions  of the 5D gauge theory 
with the cut-off $\Lambda = g v $. 
Running of the gauge couplings in `flat'  deconstruction models was studied in 
\cite{Arkani-Hamed:2001vr,Chankowski:2001hz,Csaki:2001zx}. 
 
It was noticed in \cite{Abe:2002rj} that if the quantity  $g_i v_i$ 
depends on $i$, such models correspond to 5D gauge theories 
with a  non-trivial warp factor $a(y)$. 
In particular choosing $g_i = g$  and  $v_i = v \e^i$, 
$\e^N \approx 10^{-16}$, 
we are able to describe the physics of   gauge fields 
(with a constant gauge coupling) 
living in the Randall-Sundrum background. 
One can understand it by noting, that $g_i v_i$ plays 
the role of an effective lattice spacing measured by the warped metric. 
In consequence, the scale at which we switch from 5D description 
to the UV completion  depends on the position in the group space. 
This is similar in spirit to the notion of a position dependent 
momentum cut-off introduced in \cite{Randall:2001gb}. 
 
\begin{figure} 
\begin{center}
\begin{picture}(400,200)(0,0)
\BCirc(50,150){15}
\Text(50,125)[]{\Large $G_1$}
\BCirc(150,150){15}
\Text(150,125)[]{\Large $G_2$}
\Line(65,150)(135,150)
\Text(100,165)[]{\Large $\Phi_1 = v_1$}
\DashLine(65,145)(135,145){5}
\BCirc(100,135){15}
\DashLine(100,120)(100,90){3}

\BCirc(100,75){15}
\Text(100,50)[]{\Large $G_{(2)}$}
\BCirc(200,75){15}
\Text(200,50)[]{\Large $G_3$}
\Line(115,75)(185,75)
\Text(150,90)[]{\Large $\Phi_2 = v_2$}
\DashLine(115,70)(185,70){5}
\BCirc(150,60){15}
\DashLine(150,45)(150,15){3}
\BCirc(150,0){15}
\Text(150,-25)[]{\Large $G_{(3)}$}

\DashLine(170,-15)(240,-105){10}

\BCirc(250,-125){15}
\Text(250,-150)[]{\Large $G_{(N-1)}$}
\BCirc(350,-125){15}
\Text(350,-150)[]{\Large $G_N$}
\Line(265,-125)(335,-125)
\Text(300,-100)[]{\Large $\Phi_{N-1} = v_{N-1}$}
\DashLine(265,-130)(335,-130){5}
\BCirc(300,-140){15}
\DashLine(300,-155)(300,-185){3}
\BCirc(300,-200){15}
\Text(300,-225)[]{\Large $G_{(N)}$}

\Text(175,-260)[]{\large Fig 1. Moose Diagram for Warped Gauge Theory}

\end{picture}
\end{center} 
\end{figure} 
\newpage 
 
Using eqs. (\ref{eq.gm}) one finds the connection between the parameters of 
the two theories: 
 
\bea 
\label{eq.rsd} 
\frac{1}{gv_1} & \rightarrow & \Delta y \sim \frac{1}{\Lambda}  \nn 
\frac{1}{gv_i} & \rightarrow & \Delta y \sim \frac{1}{\Lambda e^{-ky}} 
~~ ({\rm for ~~ 4D ~~ observer}) \nn 
\frac{N}{\pi g v_1} & \rightarrow & R ~, ~~ 
(\e^N \rightarrow e^{- k \pi R}) \nn 
gv_1 \log (\frac{1}{\e}) & \rightarrow & k  \nn 
%N \frac{1}{g^2} & \rightarrow & \pi R \frac{1}{g_5^2} \nn 
g v_{N-1} & \rightarrow & \Lambda e^{ - k \pi R} \nn 
\e & \rightarrow & e^{-\frac{k}{\Lambda}} ~, ~~ 
(\e \ll 1  \rightarrow  \Lambda \ll k) 
\eea 
 
The mass matrix of the gauge bosons is: 
\bea 2g^2v_1^2 \left[ \begin{array}{ccccccc} 
\\ 1 & -1 & 0 & 0 & 0 & \cdots & 0 \\ \\ 
-1 & 1 +\e^2 & -\e^2 & 0 & 0 & \cdots & 0 \\ \\ 
0 & -\e^2 & \e^2 +\e^4 & -\e^4 & 0 & \cdots & 0 \\ \\ 
\cdots & \cdots & \cdots & \cdots & \cdots & \cdots & \cdots \\ \\ 
0 & \cdots & 0 & 0 & -\e^{2(N-3)} & \e^{2(N-3)}+\e^{2(N-2)} & 
-\e^{2(N-2)} \\ \\ 
0 & \cdots & 0 & 0 & 0 & -\e^{2(N-2)} & \e^{2(N-2)} \\ \\ 
\end{array} \right] \eea 
Unfortunately, for $\e \neq 1$, it is difficult to diagonalize the mass matrix 
and  the eigenvalues must be  obtained numerically. 
However, for $\e \ll 1$, we can  diagonalize 
 the mass matrix step by step and the 
result remains trustworthy. Therefore, from now on, the case with  $\e \ll 1$ 
is considered. In such case the following simple picture holds. 
 At the scale  $v_1$ the first two groups $SU(N_c)_1 \times SU(N_c)_2$ 
are broken to the diagonal group  $SU(N_c)_{(2)}$, at the scale 
$v_2$ $SU(N_c)_{(2)} \times SU(N_c)_3$ breaks to $ SU(N_c)_{(3)}$ and so on. 
 We use the notation $SU(N_c)_i$ for the $i$-th gauge group of the original 
 product, $SU(N_c)_{(i)}$ for the unbroken gauge group after 
 the  Higgsing of $SU(N_c)_{(i-1)} \times SU(N_c)_i$ and we denote 
$g_{(i)}$ the corresponding gauge coupling. Fig. 1 shows the moose diagram. 
 
 Thus, for $\e \ll 1$, the problem is reduced  to diagonalizing  $2 \times 2$ matrices and the eigenvalues are easily  determined. 
 First we diagonalize the  part involving $A_{\mu,1}$ and $A_{\mu,2}$ and 
the  eigenvalues are ${\cal O} (gv\e)$ and ${\cal O}(gv\e^2)$. 
 One state with ${\cal O} (gv\e)$ eigenvalue is isolated and remains as 
a true eigenstate while the other with ${\cal O}(gv\e^2)$ has a 
mixing term with $A_{\mu,3}$. We can do the same procedure iteratively 
and finally we obtain the spectrum ${\cal O}(gv\e^{N-1})$ 
corresponding to the final Higgsing scale. Exact calculation shows 
that there is always a zero-mode eigenstate corresponding to the 
full diagonal subgroup. This is easily seen from the fact that the 
determinant of the original mass matrix is zero or by finding the 
zero mode eigenstate given as 
 \bea A_{\mu}^{(0)}  & = & \frac{1}{\sqrt{N}} \sum_{i=1}^{N} A_{\mu,i}, \eea 
which reflects that the zero mode of the vector field is constant 
along the extra dimension in $AdS_5$ orbifold. For other higher 
mode eigenvalues, the eigenstates are given as{\footnote{The expression 
contains a term up to ${\cal O} (\e^2)$. Here we assume $\e \ll 1$. However, 
even at ${\cal O}(\e^2)$ there can be order one or bigger than order one 
corrections due to $N$ which can be extremely large. $N\e^2$ can not be 
neglected with simple assumption $\e \ll 1$. Therefore, it is assumed also 
that $N \e^2 \ll 1$ such that the potentially dangerous correction can be 
neglected. In the realistic case explaining the gauge hierarchy, 
we need $\e^N = 10^{-16}$. The choice of $\e=\frac{1}{10}$ and $N=16$ 
satisfies the second assumption $N\e^2 = 0.16 \ll 1$.}} 
 
\bea &A_{\mu}^{(N-1)}  = & \frac{1}{\sqrt{2}} (A_{\mu,1} - A_{\mu,2}) ~~~~~~~~~~{\rm 
up~~to}~~ {\cal O}(\e^2), 
\nn &A_{\mu}^{(N-2)} = & \frac{1}{\sqrt{6}} (A_{\mu,1} + A_{\mu,2} - 2 
A_{\mu,3}) ~~~~~~~~~~{\rm up~~to}~~ {\cal 
O}(\e^2), \nn && \cdots  \nn & A_{\mu}^{(N-j)} = & 
\frac{1}{\sqrt{j(j+1)}}(A_{\mu,1} 
+ \cdots + A_{\mu,j} - j A_{\mu,j+1} ) ~~~~~~~~~~{\rm up~~to}~~ {\cal 
O}(\e^2), 
\nn && \cdots \nn &A_{\mu}^{(1)}  = & \frac{1}{\sqrt{(N-1)N}}(A_{\mu,1} + \cdots + 
A_{\mu,N-1} - (N-1) A_{\mu,N} ) ~~~~~~~~~~{\rm up~~to}~~ {\cal O}(\e^2). \eea 
 
$A_{\mu}^{(N-j)}$ has an eigenvalue of order ${\cal O}(v\e^j)$ and the 5D interpretation is clear. For higher modes ($m_n \sim v$), the corresponding wave functions are 
localized at the Planck brane. This is exactly the position 
dependent momentum cutoff for the gauge fields which Randall and 
Schwartz introduced for the calculation of the gauge coupling 
running in $AdS_5$ orbifold \cite{Randall:2001gb,Randall:2001gc}. 
Thus, the position dependent momentum 
cutoff is naturally realized in the deconstruction setup. For the 
flat extra dimension ($\e=1$), the lowest modes have fewer nodes 
since they have smaller gradients and give lighter Kaluza-Klein 
states. However, if the extra dimension is highly warped, the 
scale felt by the mode is different and the lowest excited mode is 
the one that has a variation at very near the TeV brane. 
In our analysis $\e \ll 1$ and, using eqs.  (\ref{eq.rsd}), 
this corresponds to $k \gg \Lambda$ 
which means the curvature effect is crucial. 
If the cutoff is much lower the curvature 
scale, then the net degree of freedom encountered as the energy increases is 
less than in the flat case as the Kaluza-Klein level spacing becomes larger. 
This is the momentum space version of the position dependent momentum cutoff 
and is also related to the entropy counting of black holes which is 
proportional to the area rather than the volume \cite{Randall:2002tg}. 
 
For large $N$, $\e \sim 1$ even for the Randall-Sundrum geometry  
and the second lightest mode would have two nodes 
near the TeV brane, etc. As $\e \sim 1$  corresponds to $k \ll \Lambda$,
 the cutoff is much above the curvature scale. 
Then above the curvature scale, we start to see the flat five dimensional 
physics.  
Unfortunately this case is beyond the scope of our analysis since the nice 
separation of scales is not allowed and only numerical works can give  
results. 
 
%%%%%%%%%%%%%%%%%%%%%%%%%%%%%%%%%%%%%%%%%%%%%%% 
\section{Running of gauge couplings} 
%%%%%%%%%%%%%%%%%%%%%%%%%%%%%%%%%%%%%%%%%%%%%%% 

The running behaviour  of  the gauge couplings 
in the $AdS_5$  background is an important question 
both from the theoretical and the phenomenological point of view. 
However, different regularizations show different results 
which do not exactly coincide. 
In this section we study this problem using the decontruction setup 
described in the previous section. We concentrate 
on the case where only the gauge and the link-Higgs fields 
are replicated (live in the `bulk') and the matter fields live 
on either the first or the $N$-th site, corresponding to 
 matter fields on the Planck and the TeV brane, respectively. 
 The matter fields propagating  in the `bulk' are investigated 
in  Appendix A. 
 
We find it convenient to study the running  using the bottom-up approach. 
We start  with a  gauge coupling measured at some low energy scale $M_Z$, 
$\alpha(M_Z)$. 
The running up to the scale of the lowest link-Higgs vev 
$v \epsilon^{N-1}$ proceeds as in an ordinary 4D gauge theory: 
 
\bea 
\frac{1}{\alpha(Q)}= \frac{1}{\alpha(M_Z)} 
- \frac{b_0}{2\pi}\log \left(\frac{Q}{M_Z}\right) 
& \hspace{2cm}& Q < v \epsilon^{N-1} 
\eea 
and $b_0$ is the  beta-function coefficients 
which gets contributions from the (massless) 
$SU(N_c)$ gauge fields and, 
eventually, from any matter fields which are 
effectively massless at the considered scale. 
 
At the scale $v \epsilon^{N-1}$ this coupling is identified with 
the coupling $\alpha_{(N)}$ of the diagonal  group. 
In general, the matching equation at the $n$-th threshold is: 
\beq 
\frac{1}{\alpha_{(n+1)}(v\e^n)} 
= \frac{1}{\alpha_{(n)}(v\e^n)} +  \frac{1}{\alpha_{n+1}(v\e^n)}, 
\eeq 
where $\alpha_n$ is the coupling  of the $n$-th  group 
 in the product $SU(N_c)^N$. 
For $\e \ll 1$, as explained in the previous section, 
the heavy states of gauge bosons can be considered as decoupled between the thresholds. Also, we assumed that the physical links $\Sigma_i$ are decoupled, and $G_i$ are Goldstone bosons which become longitudinal components of the heavy gauge bosons. Thus  the running is given by 
\bea 
\frac{1}{\alpha_{(n+1)}(Q)} = \frac{1}{\alpha_{(n+1)}(v\e^n)} 
- \frac{b_{(n+1)}}{2\pi}\log \left(\frac{Q}{v\e^n }\right) 
&\hspace{2cm}& v \e^{n}< Q < v \e^{n+1} 
\eea 
and the beta-function coefficients $b_{(n+1)}$ include contributions 
only from the {\it massless} $SU(N_c)$ gauge fields and from the matter 
localized  from the first up  to the  $n$-th site. 
 Since we assumed 
that all the matter fields live  either on the $N$-th or 
 on the first site we can write $b_{n} \equiv b$. 
(but still  $b \neq b_0$ if some  matter fields live 
on the $N$-th site). 
 
Putting together all these equations, 
the low-energy gauge coupling $\alpha(M_Z)$ 
depends on the high-energy couplings in the following way: 
\bea 
\frac{1}{\alpha(M_Z)}  & = & 
\frac{1}{\alpha_{1}(v)}+ \frac{1}{\alpha_{2}(v\e)} 
+\frac{1}{\alpha_{3}(v\e^2)} 
+ 
\cdots 
+\frac{1}{\alpha_{N}(v\e^{N-1})} 
\nn  
&& 
+\frac{b}{2\pi} \log \left( \frac{1}{\e^{N-1}}\right) 
+ \frac{b_0}{2\pi} \log \left(\frac{v\e^{N-1}}{M_Z}\right) 
\eea 
 
At this point there seem to be  two plausible options for 
choosing the boundary 
conditions for the gauge couplings of the product group $\alpha_n$. 
One is to choose all the gauge  couplings to be equal 
at the highest scale $v$. 
The other is to choose each gauge coupling of the product group 
so that they have the same value at  different scales, 
namely at the scale where the given group gets broken by the link-Higgs vev. 
The origin of this ambiguity comes from the question, 
 at which scale the matching between 4D and 5D degrees of freedom 
should be performed. 
The second option is motivated by  AdS/CFT correspondence 
of the Randall-Sundrum setup that the cutoff 
at different positions along the curved extra dimension 
should be different. In the folowing we investigate the consequences of 
 both approaches. 
%We can relate two boundary conditions 
%by assuming that for the first boundary condition 
%each gauge group is  conformal at the scale from $v$ to $v_i$, 
%i.e., before the gauge group is broken by the link Higgs fields. 
%Thus the second boundary condition is obtained 
%from the first one with an additional assumption. 
 
The first option consist in choosing the boundary conditions 
of the gauge couplings as 
 \bea \ia{1} |_{v} & = & \ia{2} |_{v} = \cdots \ia{N} |_{v} = \ia{}. \eea 
The RG running of each gauge  coupling 
down to the scale at which the corresponding group 
gets broken is 
\bea \frac{1}{\alpha_i(v_{i-1})} & = &  \frac{1}{\alpha_i(v)}  + 
 \frac{\tilde b}{2\pi} \log (\frac{v}{v_{i-1}}) = 
\frac{1}{\alpha} + (i-1) \frac{\tilde b}{2\pi} \log (\frac{1}{\e}). \eea 
Here the beta function coefficient $\tilde b$ contains, 
except for the contribution of the massless gauge fields, 
also the contribution from the link-Higgs degrees of freedom 
$G$ and $\Sigma_i$ (depending on the details of the scalar potential). 
>From the point of view of the 5D RS model this part is highly dependent 
on the UV completion. 
 
The low energy gauge coupling is: 
\bea 
\frac{1}{\alpha(M_Z)}  = \frac{N}{\alpha} 
+ (N-1)\frac{b}{2\pi} \log \left(\frac{1}{\e}\right) 
+ \frac{b_0}{2\pi}\log \left(\frac{v\e^{N-1}}{M_Z}\right) 
+ \frac{(N-1)(N-2)}{2} \frac{\tilde{b}}{2\pi} 
\log \left(\frac{1}{\e}\right) 
\eea 
 
For large $N$ the above formula becomes 
\bea 
\frac{1}{\alpha(M_Z)}  = \frac{N}{\alpha} + 
 \frac{N^2}{2} \frac{\tilde{b}}{2\pi} \log \left(\frac{1}{\e}\right) 
\eea 
 
Using the dictionary  given in eqs. (\ref{eq.rsd}) (in particular  
$\log (1/ \e) \ra k/\Lambda$ and  
$N \log (1/\e) = \log (1/\e^N) \ra \log(\Lambda/\mu)$)  
the corresponding equation 
 in the 5D interpretation is 
\bea \frac{1}{\alpha (\mu)} & \simeq &  \frac{\pi R}{\alpha_5} + 
\frac{\Lambda}{2 k} \frac{\tilde{b}}{2\pi} (\log (\frac{\Lambda}{\mu}))^2, \eea where we  relate $v \sim \Lambda$, $v_{N-1} \sim \mu$. Therefore, we obtain the result containing  $\log ^2 (\frac{\Lambda}{\mu})$ which looks weird. 
A similar result was obtained by  Randall and Schwartz 
before doing the  correct Greens function renormalization. 
 
The second option looks more plausible. The boundary condition is 
\bea 
\label{eq.boundary} 
\ia{1} |_{v} & = & \ia{2} |_{v_1} = \cdots 
\ia{N} |_{v_{N-1}} = \ia{}. \eea 
and the low-energy gauge couplings depends on the high-energy ones as: 
 
\bea 
\frac{1}{\alpha(M_Z)}  = \frac{N}{\alpha} 
+ (N-1)\frac{b}{2\pi} \log \left(\frac{1}{\e}\right) 
+ \frac{b_0}{2\pi}\log \left(\frac{v\e^{N-1}}{M_Z}\right) 
\eea 
 
 For large $N$,  the 5D interpretation of this equation  is: 
 
\beq 
\frac{1}{\alpha (\mu)}  =   \frac{\pi R}{\alpha_5} + 
 \frac{b}{2 \pi} \log (\frac{\Lambda}{\mu}), 
\eeq 
 
 This just shows the usual logarithmic running for the gauge 
couplings even though the deconstruction model  mimics the  gauge bosons 
propagating in 5D $AdS_5$ orbifold. Another feature which is visible here is that matter localized on the TeV brane does not contribute to the running above TeV since its contribution is included in $b_0$ but not in $b$. 
 
Another way to summarize this approach is to notice that  in deconstruction it is the coupling  $\tilde \alpha$ defined as: 
\beq 
\label{eq.alpha} 
{1 \over \tilde \alpha(Q)} \equiv {1 \over \alpha_{(n)}(Q)} 
- {1 \over  \alpha_{1}^{(i)}(v)} 
- {1 \over \alpha_{2}^{(i)}(v\e)} 
-\dots 
-{1 \over   \alpha_{n}^{(i)}(v\e^{n-1})} 
\hspace{2cm} v \e^{n-1} < Q <v\e^{n} 
\eeq 
which is logarithmically sensitive to the difference of the scales: 
\beq 
{1 \over \tilde \alpha(Q)} = - {b \over 2\pi} \ln \left ({ Q \over v} \right) 
 \hspace{2cm} 
\eeq 
and thus it should be used to study the physics of RG running in $AdS_5$. 
 
The boundary condition given in eqs. (\ref{eq.boundary}) 
is derived from 5D gauge couplings defined at UV scale 
preserving the symmetry of $AdS_5$. 
Since the actual cutoff measured by 4D observer 
located at $y$ is $\Lambda e^{-ky}$ for 5D cutoff $\Lambda$,  
the correct boundary condition necessary for 4D theory is 
\bea 
\int_0^{\pi R} dy \frac{1}{g^2(\Lambda e^{-ky})} 
& \rightarrow & \sum_{i=1}^{N-1} \frac{1}{\a_i (v_{i-1})} \nn 
\eea which is given in eqs. (\ref{eq.boundary}). 
The integrating out procedure defining the boundary condition 
(i.e., $\Lambda > M_{GUT}$) is in Fig. 2. The symmetry breaking scale 
differs for different position of $y$, and the unusual boundary condition 
is very important in checking whether the unification really occurs. 
Fig. 3 shows the region that we are considering in the paper. 
For $\Lambda < M_{GUT}$, the process to reach the boundary condition 
preserving the symmetry of $AdS_5$ involves filling up the parts 
we have integrated out which is shown in Fig. 3. 
 
\begin{figure} 
\begin{center} 
\input{fig2} 
\vspace{5mm} 
\caption{Wilsonian integrating out procedure above $M_{GUT}$ 
which keeps the symmetry of $AdS_5$} 
\vspace{5mm} 
\input{fig3} 
\vspace{5mm} 
\caption{Wilsonian integrating out procedure below $M_{GUT}$ 
which is based on 4D momentum} 
\vspace{5mm} 
\end{center} 
\end{figure} 
%%%%%%%%%%%%%%%%%%%%%%%%%%%%%%%%%%%%%%%%%5 
\subsection{5D interpretation} 
%%%%%%%%%%%%%%%%%%%%%%%%%%%%%%%%%%%%%%%%%%5 
There are two features in  warped gauge theories which are different compared 
to the flat case. For flat extra dimensions, the logarithmic 
 running becomes power law 
as we raise up the energy scale since the effective number of 
Kaluza-Klein modes circulating in the loop increases. 
More precisely, below $1/R$ there is only one mode 
contributing to the one-loop calculation. 
At energies above the compactification scale the number 
of particles  contributing to the running is determined 
by the number of KK modes with masses 
below the corresponding energy scale and for $n$ extra dimension  
it has a simple 
scaling according to energy 
\bea 
N(E) \sim E^n. 
\eea 
The differential equation is given by 
\bea 
\frac{1}{\alpha (E+\Delta E) }  & = & \frac{1}{\alpha (E)} + b N(E) \log \frac{(E+\Delta E)}{E}. 
\eea 
Integrating it from $1/R$ to $\Lambda$ one obtains 
\bea 
\frac{1}{\alpha (\Lambda)} & = & \frac{1}{\alpha (1/R)} + c_n b (\Lambda R)^n + \cdots. 
\eea 
For example,  we get a linearly cutoff dependent correction 
for the gauge theory with one extra dimension. 
But in this case the threshold correction at the cutoff scale 
is also given by $c \Lambda$. 
Unlike the case of the large logarithmic running in MSSM, 
the power law running effect is always comparable to the threshold correction. 
 
In warped gauge theories, the first difference comes from the spectrum. 
The levels of Kaluza-Klein modes are not equal spaced 
and the mass difference between adjacent one becomes 
bigger for higher KK modes. 
Therefore the effective number of KK modes below certain energy $E$ is 
\bea 
N(E) \sim \log E. 
\eea 
This softens the power running behavior by a certain amount 
but can not explain why one gets just 
a logarithmic running like in four-dimensional theory 
since the integration of the differential equation 
gives $(\log) ^2$ rather than $\log$. 
\bea 
\frac{1}{\alpha (\Lambda)} & = & \frac{1}{\alpha (1/R)} 
+ \frac{1}{2} b \left( \log (\Lambda R) \right)^2. 
\eea 
Here comes the second difference. 
In flat extra dimension, all the couplings of the KK modes 
are the same and are simply  related 
to the  higher dimensional coupling. 
In particular, the zero mode gauge coupling is related 
to  the higher dimensional gauge coupling 
by the volume factor. 
By looking at the zero mode coupling, 
we can extract higher dimensional coupling. 
However, for the warped extra dimension, 
the coupling defined at the scale above TeV is not 
a zero mode coupling but a coupling localized near the Planck brane. 
Fig. 2 shows the region in which the coupling is defined according to  
the energy scale above TeV. 
For extremely high energy scales near the Planck scale, 
the corresponding coupling is defined only at the Planck brane. 
 The discrepancy between the arguments based on effective KK modes available 
at certain energy and the result obtained in the deconstruction lies here. 
Since the coupling we are looking at does not exist in the entire interval, 
it is very crucial to know what is 
the wave function overlap of each KK mode with the corresponding region. 
The interesting observation, which can be made in the deconstruction models,  is that, 
though the effective number of KK modes is 
logarithmically increasing at high energies, 
the net contribution with the inclusion of the wave function overlap 
is always that of a single particle contribution. 
The coupling that we are interested in 
exists from $y=0$ to $y_*$ for given energy $E = \Lambda e^{-ky_*}$. 
As a specific example, let us see the energy range 
between the first and the second KK mode. 
 At this scale we study the running of the coupling defined from $y=0$ to $y_*
 = (\frac{N-1}{N}) \pi R$ and only  the zero mode and the first KK mode are light enough to contribute to the running. 
%The zero mode gives the dominant effects ($\frac{N-1}{N}$) 
%and the first KK mode gives the remaining part ($\frac{1}{N}$). 
The zero mode is constant along the extra dimension and contributes 
($\frac{N-1}{N}$) which is exactly proportional to the region 
on which the gauge coupling is defined. 
However, the first KK mode is localized 
in the region $(y_*, \pi R)$ and its contribution to the coupling in $(0,y_*)$
 is just ($\frac{1}{N}$).
Therefore, the net contribution is that of single mode.  
Even at the highest energy scale we can reach ($\Lambda$), 
the net contribution is just that of  one particle 
though all the KK modes are contributing 
to the running of the Planck brane localized gauge coupling. 
The gauge coupling is defined only near $y=0$ and 
the zero mode has a wave function overlap $\frac{1}{N}$ with the region, 
the lightest one has $\frac{1}{(N-1)N}$, the second lightest 
one has $\frac{1}{(N-2)(N-1)}$, etc., and the sum is 
$\frac{1}{N} + (-\frac{1}{N}+\frac{1}{N-1}) + (-\frac{1}{N-1}+\frac{1}{N-2}) + 
\cdots + (-\frac{1}{2} + 1) = 1$.
% This can be read off from the eigenstates. 
It is a very interesting observation captured in the deconstruction. 
The logarithmic running obtained here is in accord 
with the result of Pomarol \cite{Pomarol:2000hp} 
and Randall-Schwartz \cite{Randall:2001gb} for $\Lambda \ll k$. 
 
It should be kept in mind that the Kaluza-Klein mode analysis given here 
is just a convenient way of deriving RG equation in $AdS_5$ 
(more generally, in the curved space). 
Since  UV parameters are defined at different scales for a 4D observer, 
it is not trivial to relate the UV parameter and the experimentally measured 
IR parameter. 
In solving these kinds of problems,  
the KK analysis via deconstruction is very useful and shows the physics 
transparently. 
However, this does not mean that we can derive physical spectrum above TeV. 
 
%%%%%%%%%%%%%%%%%%%%%%%%%%%%%%%%%%%%%%%%%%%%%%%%5 
\subsection{Comment on the continuum limit} 
%%%%%%%%%%%%%%%%%%%%%%%%%%%%%%%%%%%%%%%%%%%%%%%%%%% 
 
 Since the low energy gauge coupling is given by that 
of the diagonal subgroup, it is not possible to raise $N$ to be 
arbitrary large while keeping the gauge groups above the 
deconstruction scale weakly coupled. In the case of flat extra 
dimensions, \bea \frac{1}{\alpha} & = & \frac{N}{\alpha_i}, \eea 
and for the low energy coupling $\alpha \sim 1/20$, we get the 
maximum number of $N$ to be $240$ with $\alpha_i \sim 4\pi$. 
Similar restriction applies here. If we restrict our analysis to 
warped gauge theory with known hierarchy, the restriction on $N$ 
is related to the upper bound on the cutoff compared to the 
curvature scale. With the notation given above, \bea \e^N & \sim & 
10^{-16}, \nn  \e & = & e^{-\frac{k}{\Lambda}},  \eea and we get 
\bea \frac{k}{\Lambda} & \ge & \frac{1}{6}. \eea Thus $\Lambda = 6 
k$ is the maximum cutoff we can reach for the bulk Randall-Sundrum 
setup {\footnote{The bounds on $\Lambda$ might be stronger if we 
consider $\Lambda > k$. In the derivation we used flat space 
constraint on $\alpha$ which is very similar for $\e \ll 1$. 
However, for $\e \sim 1$, power law correction appears and the 
restricted value of $N$ might be smaller than 240.}. This is 
enough to study the interesting physics since $\Lambda$ is high 
enough. 
 
However, as we increase $N$, the neglected corrections 
become comparable to the log correction. 
In the derivation of the running equation for the gauge couplings, 
it is assumed that $\e \ll 1$ such that $\log (\frac{1}{\e}) \gg {\cal O}(1)$ 
at each matching scale. 
Once $\e$ becomes close to one, the threshold correction can not be neglected. 
At the same time, the accumulated effect gets bigger 
as $N$ increases. 
The power law correction $N \e^2$ becomes of order one 
for $N \sim 30$ with $\e \sim 1/5$. 
This is when $\Lambda \sim 1.6k$. 
For $\e \sim 1/5$, the logarithmic correction is already the same order 
as order one threshold correction. 
Therefore, the logarithmic correction derived here 
is spoiled by the power law correction unless $\e \ll 1$. 
Furthermore, there is a threshold correction at the curvature scale 
which might not be universal when the GUT breaking scale ($\Lambda$) 
is above the curvature scale. 
 
For the unified theory living on a (flat) higher dimension, 
the threshold correction is not negligible 
and we can not make a clean prediction 
based on the low energy gauge couplings \cite{Arkani-Hamed:2001vr}. 
The same problem appears if we raise our cutoff 
(the GUT breaking scale) $\Lambda$ 
to be higher than the curvature scale $k$. 
 
%%%%%%%%%%%%%%%%%%%%%%%%%%%%%%%%%%%%%%%%%%%%%%%%%%%%%%%%%%%%%%%%% 
\section{Conclusions} 
%%%%%%%%%%%%%%%%%%%%%%%%%%%%%%%%%%%%%%%%%%%%%%%%%%%%%%%%%%%%%% 
In a gauge theory in the $AdS_5$ background, unlike in the flat case, 
 the gauge coupling runs {\it logarithmically}. 
This observation may shed light on the 
possibility of having a unified theory in the Randall-Sundrum framework. 
In this paper we studied the running of gauge couplings 
 in four-dimensional deconstruction models which mimic warped gauge theories. 
Deconstruction allows us to understand some properties of 
higher dimensional gauge theories with the help of 
the standard tools of  four-dimensional gauge theories. 
 
When the scale difference for different link-Higgs fields is large 
enough, $\e \ll 1$, which corresponds to the cutoff of the 5D theory 
 lower than the curvature scale, the running is purely logarithmic. 
It might be more natural to keep the cut-off above the curvature scale 
so that  one could define the theory at high-energies 
in an effectively flat background. 
But once we do this, 
the cutoff dependent threshold correction and the 
curvature dependent threshold correction ruin 
the predictability for the gauge unification 
since these unknown corrections are comparable to the calculable part. 
This gives us some insight on the features 
that a highly predictable GUT theory should possess. 
 
Once the GUT breaking scale ($M_{GUT}$) is below the curvature scale, 
then above $M_{GUT}$ three gauge couplings get common corrections, 
and the different correction start to appear below $M_{GUT}$. 
Then the result obtained in the paper gives a high precision prediction 
on the possibility of the unification regardless of the physics 
above $M_{GUT}$ which is common to three gauge couplings. 
 
Constructing realistic unification model of the warped gauge theory 
which can avoid proton decay remains as a future work. 
 
\section*{acknowledgement} 
 
HD thanks K. Choi, K. Dienes and E. Poppitz for discussions 
and thanks M. Shifman for the criticism on the validity of the setup. 
The work of  AF was partially supported  by the EC Contract 
HPRN-CT-2000-00148 for years 2000-2004. AF acknowledges 
hospitability of KIAS in Seoul, where part of this work was completed. 
 
 %%%%%%%%%%%%%%%%%%%%%%%%%%%%%%%%%%%%%%%%%%%%%%%%%%%%%%%%%%% 
\section{Appendix A: Bulk fields} 
%%%%%%%%%%%%%%%%%%%%%%%%%%%%%%%%%%%%%%%%%%%%%%%%%%%%%%%%%%% 
In this appendix we investigate the effect of a replicated  scalar 
field on the running of gauge couplings in the deconstruction set-up. 
 Following  the analysis of Section III we will keep track 
of the evolution of   the coupling $\tilde \alpha(Q)$ defined in 
eq. (\ref{eq.alpha}). 
 
 A complex 5D bulk field $h(x^\mu, y)$ in the fundamental representation 
  is accounted for in decostruction by putting at each site 
 a complex field $h_i(x^\mu)$ in the fundamental representation of 
 the $i$-th group. Moreover, scalars living on neighbouring sites 
should communicate via the link-Higgs fields. In a non-supersymmetric set-up, 
 the most general set of such couplings is very complicated 
so in the following we restrict ourselves  to  just 
one specific example when kinetic and interaction  terms  are: 
\beq 
\label{eq.ds} 
\sum_{i=1}^N \left ( \pa_\mu h_i^\dagger  \pa_\mu h_i 
 - \sum_{i=1}^{N-1} |g \Phi_i h_{i+1} - m_i h_i|^2   \right) 
\eeq 
 
In the continuum limit this lagrangian corresponds to: 
\beq 
\label{eq.fs} 
\L = - \int dy \left ( \pa_\mu h^\dagger  \pa_\mu h - 
 a^2(y)|\pa_5 h - M \epsilon(x_5) h|^2 \right) 
\eeq 
 
where $g_i v_i \sim { a(y_i) \over d 
y}$ and $m_i - g_i v_i \sim a(y_i) M(y_i)$. 
Thus for a choice $v_i = v \e^i$, $m_i = x g v_i$ 
our deconstruction  lagrangian (\ref{eq.ds})  mimics the 5D bulk scalar 
in the RS background (appropriately rescaled so as to get 
the kanonical 4D kinetic term) with  a constant kink-mass term 
\beq 
\label{eq.mm} 
M \sim  (x-1)g v 
\eeq 
Note that both lagrangian of eqs. (\ref{eq.fs}) and (\ref{eq.ds}) 
 yield a massless mode. In the 5D case its profile is given by 
$h^{(0)} = \exp ( M |y|)$ while in the deconstruction 
it is $h^{(0)}_i = x^i$. Since the matching (\ref{eq.mm}) yields 
 $x \sim 1+ M/(g v) \sim 1 + M \Delta y$ we see 
we can reliably mimic scalar with a kink-mass not greater than 
$(\Delta y)^{-1}$. 
 
We are now ready to investigate the effect of such replicated scalar field 
on  the running. We consider seperately three cases: 
\be 
\item  $ x = 1 $ (vanishing bulk mass term) \\ 
Above $v \e $, 
${ 1 \over \tilde \alpha(Q)}$   is defined as 
 ${1 \over \alpha_{(1)}(Q)} -{1 \over \alpha_1(v)}$  thus 
  only $h_1$ (which is effectively massless at this scale) 
 contribute to its running. 
 
 At $v\e$ two groups are effectively broken and both 
 $h_1$ and $h_2$ can  contribute to the running of 
 ${1 \over \tilde \alpha(Q)} =  {1 \over \alpha_{(2)}(Q)} -{1 \over  \alpha_1(v)} - {1 \over \alpha_2(v\e)}$. 
The relevant mass terms are: 
\beq 
\L = - (g v \e)^2 |h_1 - h_2|^2 
\eeq 
Diagonalizing the mass matrix, we find that (for $g \sim 1$) 
one eigenvalue is $2 v \e $ thus the corresponding state 
 is decoupled below $v \e$. The other eigenvalue is zero 
 (or more precisely, less than $v \e^2$), 
 hence we conclude that down to $v\e^2$ 
 the bulk scalar gives the replicated of 
{\it one massless scalar} to the running of $\tilde \alpha(Q)$. 
 
Repeating this analysis at the consecutive thresholds we always 
find the same qualitative picture: there is one combination of 
the scalars $h_p$  which is effectively massless and the remaining 
states are decoupled. Thus, in  deconstruction picture, for a replicated 
 scalar without bulk mass term  only the zero-mode contributes to the 
 running of $\alpha$ and the running is given by: 
\beq 
\label{eq.ar} 
{1 \over  \tilde \alpha(Q)} = - {b_S \over 2\pi} \log \left ({ Q \over v} \right) 
 \hspace{2cm} 
\eeq 
where $b_S$ is the beta-function coefficient of a massless scalar 
in the fundamental representation. 
 
\item $x<1$, (negative kink mass)\\ 
In this case, the zero mode is localized towards the first site and thus it 
should contribute to the evolution of the coupling from high down 
to low scales, just like a field on the Planck brane. 
In fact, the analysis is very similar to the case $x=1$ 
 described in the preceding paragraphs. 
At each threshold there is one massless state which contributes 
 to the running and the remaining states can be considered  decoupled. 
 Thus, also for $x<1$,  the running is given be eq. (\ref{eq.ar}). 
 
\item $x>1$, (positive kink mass)\\ 
In this case, the zero mode is localized towards the $N$-th 
site and  thus it is not necessarily seen at high-energies. 
 The analysis is more complicated and, in general, 
 an exact formula for the running cannot be found. 
 
Above $v \e $, only  $h_1$ can contribute to the running of $\alpha$. 
 Its mass  is $x v \e$ thus it decouples  before reaching the $v\e$ threshold. 
 Thus, the running is given by: 
\beq 
\label{eq.ar2} 
{1 \over \alpha_{(1)}(v\e)} ={1 \over  \alpha_1(v)} - {b_S \over 2\pi} 
\log \left ({  x \e v  \over v} \right) 
 \hspace{2cm} 
\eeq 
 
 In the range  $(v\e, v\e^2)$  $h_1$ and $h_2$ can  contribute to 
 the running of $\tilde \alpha(Q)$ and  their mass terms are: 
\beq 
\L = - (g v \e)^2  |h_2 - x h_1|^2 - (g x v \e^2)^2 |h_2|^2 
\eeq 
The greater eigenvalue is approximately $(x^2+1)v g \e$ 
 and the corresponding state is decoupled. 
  The lower eigenvalue is approximately $(x^2-1)g v \e^2$ 
 and for $x > \rt$ this state decoples before reaching $v\e^2$. 
 Thus the running is: 
\bea 
{1 \over \alpha_{(2)}(v\e^2)} & = & {1 \over \alpha_{(2)}(v\e)}  - {b_S \over 2\pi} 
\log \left ({  (x^2-1) \e^2 v  \over v \e} \right) \nn 
& = & 
{1 \over \alpha_1(v)}+ {1 \over \alpha_2(v\e)} - {b_S \over 2\pi} 
\log \left ({\e^2} \right) - {b_S \over 2\pi}\left (x (x^2-1) \right) 
\eea 
 
Repeating the same analysis for consecutive thresholds we obtain: 
\beq 
{1 \over \tilde \alpha(Q)} = 
- {b_S \over 2\pi} \log \left ({Q \over v} \right) 
 - {b_S \over 2\pi} \log \left (f({M\over v}, Q) \right) 
\eeq 
 
Compared to the two previous cases there appears a correction 
 depending on some function $f$ of the effective kink mass and the RG scale $Q$. 
 
\ee 
 
\section{Appendix B : Three-site Model} 
 
It is meaningful to look at the simplest model showing interesting physics. 
Since it is not possible to analyze N-site model in general 
exactly without numerical analysis, cascading procedure is used to 
study warped gauge theory. This requires enough amount of scale 
differences for VEVs of different sites and the flat limit can not 
be taken continuously within this analysis. 
For three sites, it is possible to 
diagonalize $3\times 3$ matrix exactly and this allows us to see 
both the flat and warped gauge theory limit at the same time by taking 
the parameters appropriately. Therefore in this appendix we study 
three site model{\footnote{For two site model, 
it has only one scale for the VEV of link Higgs 
which always gives flat space geometry 
and warped geometry can not be seen.}}. 
>From the general expression, 
we get \bea \left( \begin{array}{ccc} 1 & -1 & 0 \\ -1 & 1+ \e^2 & 
-\e^2 \\ 0 & -\e^2 & \e^2 \end{array} \right) \eea for $N=3$. 
Three eigenvalues are \bea m_0^2 & = & 0, \\ m_1^2 & = & (1+\e^2) 
- \sqrt{1-\e^2+\e^4}, \\ m_2^2 & = & (1+\e^2) + 
\sqrt{1-\e^2+\e^4}. \eea Corresponding eigenstates are \bea 
A^{(0)} & = & \frac{1}{\sqrt{3}} ( A^1 + A^2 + A^3), \\ A^{(1)} & 
= & C(\e) ( A^1 + (-\e^2 + \sqrt{1-\e^2+\e^4}) A^2 + (-1+\e^2 - 
\sqrt{1-\e^2+\e^4}) A^3 ), \\ A^{(2)} & = & C^{\prime} (\e) ( A^1 + (-\e^2 - 
\sqrt{1-\e^2+\e^4}) A^2 + (-1+\e^2 + \sqrt{1-\e^2+\e^4}) A^3 ), 
\eea 
where $C(\e)$ and $C^{\prime}(\e)$ are the normalization coefficients. 
 
Two interesting limits are 
\begin{enumerate} 
 
\item Gauge theory in flat space($\e = 1$) \\ 
The eigenvalues are $3$, $1$ and $0$ and the corresponding 
eigenstates are $(1,-2,1)$, $(1,0,-1)$ and $(1,1,1)$ respectively. 
This shows the dependence of KK mass on the variation of the 
coefficients. The heaviest mode has the biggest variation along 
the index (corresponding the coordinate along the extra dimension 
in the geometric interpretation). 
 
\item Slightly warped gauge theory ($\e^2 = 1 - \delta$, $0 < \delta \ll 1$)\\ 
This corresponds to the case in which $k \ll \Lambda$. 
The eigenvalues are $3-\frac{3\delta}{2}$,$1-\frac{\delta}{2}$ and $0$ 
and the eigenstates are $(1+\frac{3\delta}{4}, -2, 1-\frac{3\delta}{4})$, 
$(1-\frac{\delta}{4}, \frac{\delta}{2} , -1-\frac{\delta}{4})$ and 
$(1,1,1)$ respectively. The eigenvalues are slightly smaller than the flat 
limit and the eigenstates start to show the tendency that the first excited 
mode has a more or less similar value for $A_1$ and $A_2$ while $A_3$ becomes 
bigger. For the heavest mode, $A_3$ part decreases and $A_1$ and $A_2$ becomes 
to be a similar magnitude with opposite sign. 
 
\item warped gauge theory($\e \ll 1$) \\ 
The eigenvalues are $2+\frac{\e^2}{2}$, $\frac{3\e^2}{2}$ and $0$ 
and the eigenstates are $(1,-1,0)$, $(1,1,-2)$ and $(1,1,1)$ 
respectively. The wave function shows the sharp peak at the 
position whose local cutoff is similar to the Kaluza-Klein mass. 
The zero mode has a constant wave function. The first excited mode 
has a peak at the third site. The second (highest) excited mode 
has a peak at the second site. 
 
\end{enumerate}

\end{document}